\documentclass[preprint2]{aastex}

\shorttitle{Accuracy of the exposure time calculators}
\shortauthors{Li Causi, De Marchi, Paresce}

\usepackage{natbib}

\begin{document}

\title{On the accuracy of the S/N estimates obtained with the exposure time
calculator of the Wide Field Planetary Camera 2 on board the Hubble Space Telescope}

\author{Gianluca Li Causi}
\affil{Osservatorio Astronomico di Roma, Via Frascati 33, 00040,
Monteporzio Catone (RM), Italy}
\email{licausi@coma.mporzio.astro.it}

\author{Guido De Marchi\altaffilmark{1}}
\affil{European Space Agency,
Space Telescope Science Institute, 3700 San Martin Drive, Baltimore, MD
21218, USA}
\email{demarchi@stsci.edu}

\and

\author{Francesco Paresce}
\affil{ European Southern Observatory, Karl-Schwarzschild-Str. 2,
D-85748 Garching bei M\"unchen, Germany }
\email{fparesce@eso.org}

\altaffiltext{1}{Affiliated with the Research and Science Support
Department of ESA}

\begin{abstract}

We have studied the accuracy and reliability of the exposure time calculator (ETC) of the Wide Field Planetary Camera 2 (WFPC2) on board
the Hubble Space Telescope (HST) with the objective of determining how well it represents actual observations and, therefore, how much confidence can be invested in it and in similar software tools. We have found, for example, that the ETC gives, in certain circumstances, very optimistic values for the signal-to-noise ratio (SNR) of point sources. These values overestimate by up to a factor of 2 the HST performance when simulations are needed to plan deep imaging observations, thus bearing serious implications on observing time allocation. For this particular case, we calculate the corrective factors to compute the appropriate SNR and detection limits and we show how these corrections vary with field crowding and sky background. We also compare the ETC of the WFPC2 with a more general ETC tool, which takes into account the real effects of pixel size and charge diffusion. Our analysis indicates that similar problems may afflict other ETCs in general showing the limits to which they are bound and the caution with which their results must be taken.

\end{abstract}


\section{Introduction}

ETCs play an important role in modern instrument use as they allow
observers to determine how to carry out specific investigations and,
especially, to predict the amount of time these will require. Since the
time needed for the various programmes is a very sensitive issue in the
allocation process for most modern high visibility ground and
space-based facilities, the accuracy of these simulators must be well
understood both by the observers and the time allocation committees
that must rely on their results for a fair and scientifically effective
distribution of the available time. In this context, unfortunately,
besides the documentation accompanying the software tools, there is 
practically no published information on the reliability of existing 
ETCs of imaging cameras.

The WFPC2 has been so far the principal instrument on board the HST and
it is expected to be of extreme utility to image parallel fields even
now that the Advanced Camera for Surveys (ACS) is installed on the HST.
ETC software utilities are available on the internet site of the STScI
which simulate analytically the photometry for a given target for each
HST instrument.  The accuracy of these programmes plays a fundamental
role in the planning of observations, in particular when extremely deep
imaging is required and whenever the performances of two different
instruments have to be compared.

While performing simulations for an HST proposal for the WFPC2 and the
ACS, in which high accuracy was needed in order to evaluate the
limiting magnitudes for deep observations of a globular cluster, we
found substantial differences between the WFPC2 ETC results and real
photometry obtained on archival images. We found similar differences
also in archival non crowded fields, so that we decided to analyse the
problem by directly comparing the ETC predictions with our photometry
in various circumstances and here we show the results and the way in
which they depend on field crowding. We also compare our photometry
with the result of the recently published ``ETC++'' software
\citep{ber01}, whose calculations are based on statistical analysis
tools and take into account the real effects of the pixel size and
charge diffusion.

\section{The WFPC2 ETC: comparison with real point--source photometry}

The WFPC2 ETC computes the expected SNR of a point source from its
input parameters, namely: the magnitude of the star in a given spectral
band, the spectral type, the filter to use, the channel of the detector
(PC1 or WF2, WF3, WF4), the analogue to digital gain, the position of the
star on the pixel (centre or corner), the exposure time of the whole
observation (i.e. the sum of all the exposure frames) and the sky
coordinates of the target \citep{bir96}. As of late, the option to
manually select a specific value of the sky brightness has been added
\citep{bir01}.

First, the programme computes the source count rate, assuming a
blackbody spectrum, if the user has not specified it, and multiplies it
by the response curves of the detector and filter. Then, the programme
takes into account the various noise sources, including photon noise,
read noise, dark noise and sky noise. The latter depends on the target
position on the sky, with the sky brightening by about one magnitude
from the ecliptic pole to the ecliptic plane. The programme uses the
values from Table 6.4 in the WFPC2 Instrument Handbook
\citep{biretal01} to compute the sky count rate per pixel and hence its
photon noise. The contribution of the total noise to the photometry of
a star depends upon the number of pixels in the point spread function
(PSF) and how these pixels are weighted during data reduction. The
WFPC2 ETC assumes that the data reduction employs PSF fitting
photometry, so that it weights the pixels in proportion to their
intensity, which maximises the SNR. The multiple read errors for a
``cosmic--ray splitted'' (CR-split) image, i.e. an image composed by
many shorter frames, is then computed for a set of default splitting
values and the corresponding SNR is also given in the ETC result page.

In order to quantify the possible WFPC2 ETC deviations from real
photometry, we performed accurate aperture photometry (using the
DAOPhot package) on both crowded and non crowded archival fields. The
average image used in our analysis was computed after aligning the
individual frames in the dithering pattern and removing cosmic ray
hits. A custom programme was used, which computes the offsets of the
frames by measuring the mean displacement of the centroid of some
reference stars. The task then registers all the images to the first
one, creates a mask of the CR-contaminated pixels, by means of an
iterative sigma clipping  routine with respect to the median value of
the corresponding pixels in all frames, and finally computes the mean
image by averaging the corresponding pixel on all the images if not
included in the CR-mask. The CR-corrupted pixel in the original
un-shifted frames are also replaced by the value of the corresponding
pixel in the mean image in order to allow us to perform photometry on
both the combined and on the individual frames.  Our instrumental
magnitudes were then transformed to the Johnson/Cousin UBVRI system by 
following the prescription of \citep{holt95} --- specifically,
their equation\,8, which also takes account of the colour correction by
means of the coefficients in Table~7 therein --- and by making an optimal
choice of the aperture radius for each star, so as to minimise the
associated photometric error.

\begin{figure}
\plotone{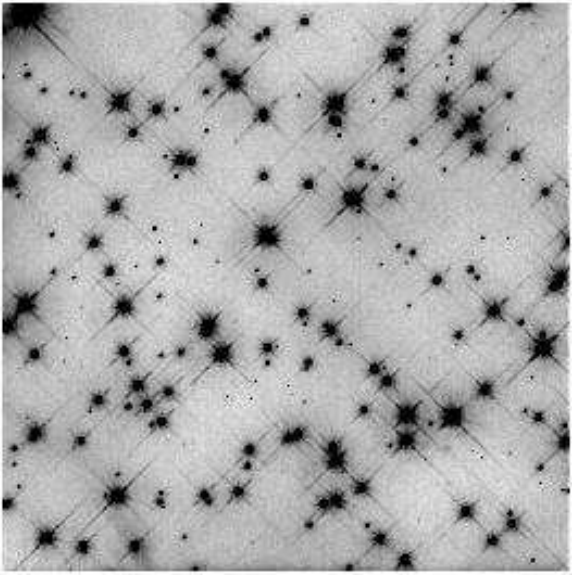}
\caption{ Negative image of a crowded field in the globular cluster M4
obtained with the PC channel of the WFPC2 through the F555W filter
(\citet{rich97}). \label{fig1}}
\end{figure}

For the crowded field, we have used the images of the Galactic globular
cluster $M\,4$ (HST proposal 5461), obtained in the F555W and F814W
filters.  These are deep images centered in a region at one core radius
from the centre of a dense globular cluster and should be
representative of the cases in which the field  observed is filled
with  a multitude of very bright and saturated stars ($V \leq 16$),
whose haloes overlap each other and cover a significant fraction of the
frame (Figure~\ref{fig1}).

\begin{figure}
\plotone{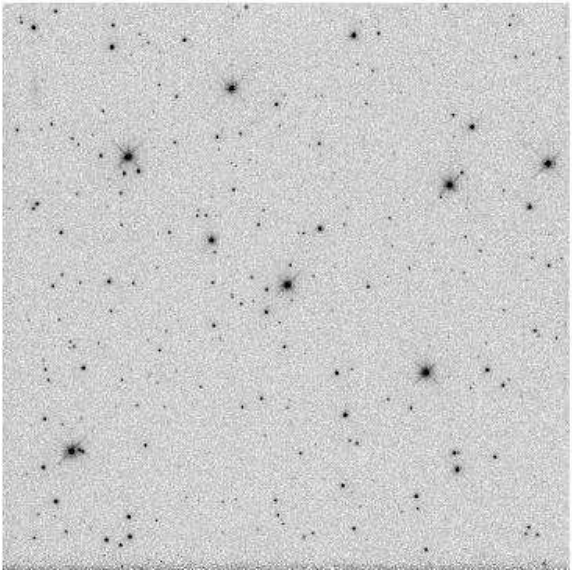}
\caption{ Negative image of a sparse field obtained with the PC (\citet{iba98}).
\label{fig2}}
\end{figure}

Images of the field of ARP\,2 taken from HST proposal 6701, also
obtained through F555W and F814W filters, were used as representative
of a sparsely filled region in which the field is populated with
faint stars with no appreciable overlapping haloes
(Figure~\ref{fig2}).

\subsection{How we measured the SNR}

Aperture photometry was performed on both series of images (i.e.
crowded and non crowded), with the following parameters. The flux of
the object was sampled within an aperture of radius $r_0$, which is
varied in steps of $0.5$ pixel. The background is sampled within an
annulus drawn from an inner radius $r_1=r_0 + 1$ pixel to an outer
radius $r_2$, with an annulus width which is varied from $3$ pixel up to
$20$ pixel. As is discussed later in this section, an adjustable
aperture radius and annulus size allow us to maximise the SNR, by
limiting the noise generated by the contamination of the neighbouring
objects. Moreover, the background is always estimated by taking the
mode, rather than the mean or median, of the pixel distribution within
the annulus. Appropriate aperture corrections were applied, which were
directly measured from the most isolated non saturated stars in the
field. A direct comparison with the encircled energy curve for
the WFPC2 PSF \citep{biretal01} shows a perfect match, thus proving
that the growth curves that we measured are reliable.

The DAOPhot task, used with the optimal aperture radius $r_0$ and the
radii $r_1$ and $r_2$ for the sky annulus, gives the best estimate of
both the magnitude and the associated error $\sigma_m$, from which
we compute the SNR by using the equation:
 
\begin{equation}
\label{eq1}
{\rm SNR^D} = \frac{1}{{\rm e}^{-\sigma_m/1.08574} -1}
\end{equation}

that comes from inverting Pogson's relation $\Delta m = - 2.5 \, {\rm
log}((F + \Delta F)/F)$, where the numerical constant $1.08574$ is
equal to $2.5 \, {\rm ln}(10)$.  Hereafter, the acronym ${\rm SNR^D}$
indicates the SNR estimated on the basis of the photometric error given
by DAOPhot.

As an independent check, we have computed the SNR as indicated in
equation 6.7 of the WFPC2 Handbook \citep{biretal01} which, in the
practical case of observed quantities, becomes:

\begin{equation}
\label{eq1a} 
{\rm SNR^H } = \frac{F \cdot G^{1/2} \cdot N^{1/2}}
{\sqrt{F + (S + N_R^2 / G) \cdot \pi \cdot r_0^2 + N_S}}
\end{equation}

where $r_0$ is the optimal aperture radius used by DAOPhot, $N$ is the number of frames combined together, $N_R$ the read-out noise (in units of electrons) of each specific CCD, $S$ the average background per pixel inside the annulus from $r_1$ to $r_2$, in units of DN, $F$ is the flux within the aperture of radius $r_0$ after subtraction of the background contribution
$S\cdot \pi \cdot r_0^2$, in units of DN, and $G$ is the effective gain factor, i.e. the CCD gain times the number of frames averaged togheter.  Finally, $N_S$ is a small (although non negligible) contribution to the error affecting the estimate of the background level which takes on the form:

\begin{equation}
\label{1b}
N_S = \frac{(S + N_R^2 / G) \cdot \pi \cdot r_0^4}
{r_2^2 - r_1^2}
\end{equation}

The computation of ${\rm SNR^H}$ makes no use of the error estimate on
the magnitude or flux provided by DAOPhot, so it is reassuring to find
that ${\rm SNR^H}$ is in excellent agreement with ${\rm SNR^D}$. This,
however, only happens if we use an adaptive choice for aperture radius
and for the background annulus, as explained above. In fact, if we
select a fixed radius and annulus size in a crowded environment, the
contamination due to neighbouring stars alters the statistics of the
sky within the annulus and we always find ${\rm SNR^H} > {\rm SNR^D}$.
This is precisely the reason that made \citet{demetal93} conclude
that core aperture photometry, i.e. source and sky measurement
conducted as close to the source as possible, as well as the use of the
mode for the background are most advisable in crowded environments.

\subsection{How the ETC expects the SNR to be measured}

In light of the consistency between ${\rm SNR^D}$ and ${\rm SNR^H}$ and
since the latter stems directly from equation 6.7 of the WFPC2
Handbook, on which the WFPC2 ETC is also based, we can now proceed and
compare our measured  ${\rm SNR^D}$ with the ETC predictions. Before
doing so, however, we must make sure that the way in which we measure
the SNR (i.e. ${\rm SNR^D}$) is consistent with the way in which  the
ETC software expects users to carry out the photometry. In fact, the
latter assumes that the data reduction process employ PSF fitting
photometry, i.e. that optimal weighting be assigned to each pixel in
proportion to its intensity in the PSF. As discussed above, however, we
have used aperture photometry to determine ${\rm SNR^D}$. The WFPC2 ETC
instructions would indeed offer a correction to apply to the ideal PSF
fitting case ${\rm SNR^P}$ (we call it ``ETC optimal SNR'') in order to
convert it to the equivalent SNR that would be obtained with canonical
aperture photometry ${\rm SNR^A}$ (``ETC aperture SNR''). Following the
WFPC2 ETC instructions in \citep{bir96}, we have:

\begin{equation}
\label{eq2}
{\rm SNR^A}= {\rm SNR^P}\cdot \frac{K}{r_0} 
\end{equation}

where $K=0.11$ for the PC camera and $K=0.17$ for the WF chips,
particularly valid when the aperture radius is $r_0>2.5$ pixel for the
PC and $r_0>1.8$ pixel in the case of the WF.

\begin{figure}
\plotone{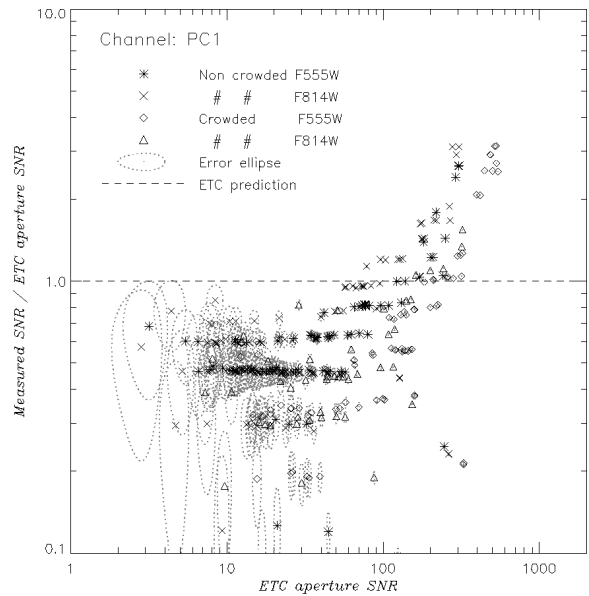}
\caption{ The ratio between the SNR measured in crowded and sparse
fields (SNR$^{\rm D}$ in the text) and the WFPC2 ETC prediction for
aperture photometry (SNR$^{\rm A}$ in the text) is shown for the F555W
and F814W filters.
 \label{fig3}}
\end{figure}

Since we determined ${\rm SNR^D}$ by using aperture photometry, it
would seem that we need to take into account the correction given by
Equation\,4. We show, however, that this correction is not necessary
thanks to the adaptive method that we used for photometry. In
Figure\,\ref{fig3} we plot, for the PC chip, the measured ${\rm SNR^D}$
against the prediction of the ETC for the aperture photometry case,
i.e.  ${\rm SNR^A}$. We should like to clarify here how
Figure\,\ref{fig3} and others of the same type in the following were
built. After having measured the calibrated magnitude of a star in the
images, we folded the latter value through the WFPC2 ETC so as to
calculate the estimated SNR for an object of that brightness and for
the exposure time and CR-SPLIT pattern corresponding to those of the
actual combined image.  For this and all the other figures in this
paper, unless otherwise specified, we used the ``average sky'' option
for the sky brightness setting as allowed by the new WFPC2 ETC Version
3.0.

\begin{figure}
\plotone{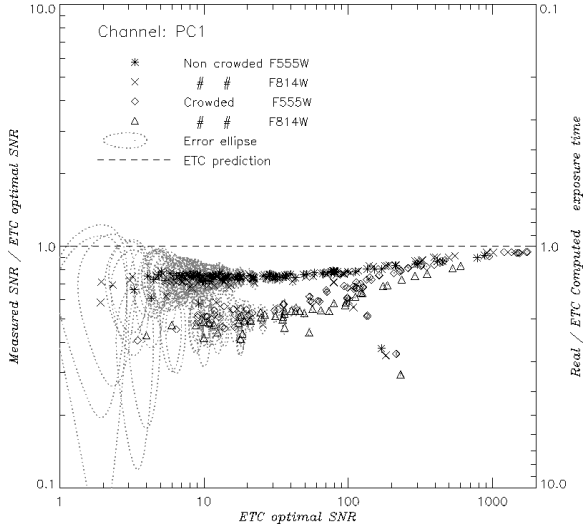}
\caption{ The ratio between the SNR measured in crowded and sparse
fields (SNR$^{\rm D}$ in the text) and the WFPC2 ETC prediction for
PSF-fitting photometry (SNR$^{\rm P}$ in the text) is shown for the
F555W and F814W filters.  The right hand side axis applies to the low
SNR regime ($\lesssim 50$) and indicates the amount of the time
estimation error, i.e. the ratio between the actual exposure time (Equation\,7 in the text) and that estimated by the ETC, for a given SNR in the abscissa.
\label{fig4}}
\end{figure}

We can see from Figure~\ref{fig3} that the prediction of the ETC for
aperture photometry (${\rm SNR^A}$) are over-estimated for faint stars
and under-estimated for bright objects with respect to the measured
values for both the sparse and the crowded field.  Figure~\ref{fig4} is
the analogue of Figure~\ref{fig3} but here the reference is the ETC
optimal SNR, ${\rm SNR^P}$, i.e. without any correction for aperture
photometry.  As one can easily see, the ETC in this case always
overestimates the value of the SNR with respect to the measured ones by
up to $\sim 100\,\%$ for the fainter stars.  As the right hand side
axis shows, such a mismatch of the SNR corresponds to a time estimation
error of the same amount (see Equation\,7 ahead), i.e. the ETC appears to underestimate the exposure time actually needed to achieve a given SNR.

A closer look at Figure~\ref{fig4}, however, reveals that the scatter
of the representative points on the plot is smaller when our
measurements are compared with SNR$^{\rm P}$ than for SNR$^{\rm A}$ and
that the overall behaviour is closer to the ETC prediction at any
magnitude. This is a consequence of our optimised aperture and annulus
photometry closely approaching PSF fitting. In light of these results,
in the following we ignore the correction for aperture photometry given
by equation\,4 and compare our measurements directly with the ETC
optimal SNR, i.e. SNR$^{\rm P}$.

\subsection{How the predicted SNR compares with the observed one}

Figures~\ref{fig3} and~\ref{fig4} clearly witness the dependence of the
actual SNR upon the level of field crowding and, at the same time, its
independence of the filter used. In principle, one could question the
validity of our latest assumption, i.e. that of ignoring the
correction to be applied to the SNR measured with aperture photometry.
In fact, in a crowded field, PSF fitting photometry is expected to give
better results. We have, therefore, attempted a direct comparison
between the predictions of the ETC and the results of PSF fitting
photometry. Rather than carrying out the reduction ourselves, we have
utilised one of the finest examples of photometric work carried out on
these very M\,4 data by \citet{rich97}, who employed very accurate
ALLFRAME photometry as described in detail in \citet{iba99}.  In their
paper, these authors measure the magnitude of each star from the
individual frames in the dithering stack and compute the combined
magnitudes as the weighted average of the corresponding fluxes, the
error on them, $\sigma_m^P$, being related to the flux scatter amongst
the frames.

\begin{figure}
\plotone{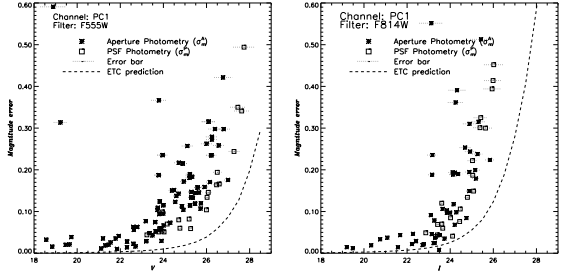}
\caption{ Measured magnitude error from PSF-fitting photometry of
\citet{iba99} ($\sigma_m^P$) and from our optimized aperture photometry
($\sigma_m^A$), as a function of the magnitude, compared with the WFPC2
ETC prediction, for the crowded field and F555W filter.
\label{fig5} }
\end{figure}

In order to make a reliable comparison with our results, we have
performed, in a similar way, optimised aperture photometry on the
individual frames (i.e. the original, not yet aligned images, in which
CR-hits had been removed as described above). The measured fluxes were
averaged with a weight inversely proportional to the DAOPhot estimated
uncertainty after rescaling for the flux ratio. Our final magnitude
errors, $\sigma_m^A$, are thus derived from the standard deviation of
the fluxes, divided by the square root of the number of images
combined.  Figure~\ref{fig5} displays the comparison between
$\sigma_m^P$, $\sigma_m^A$ and the ETC prediction, showing that the two
photometric uncertainties overlap each other, while the ETC largely
overestimates the precision that can be attained with PSF fitting
photometry, even by one of the most experienced teams.

\begin{figure}
\plotone{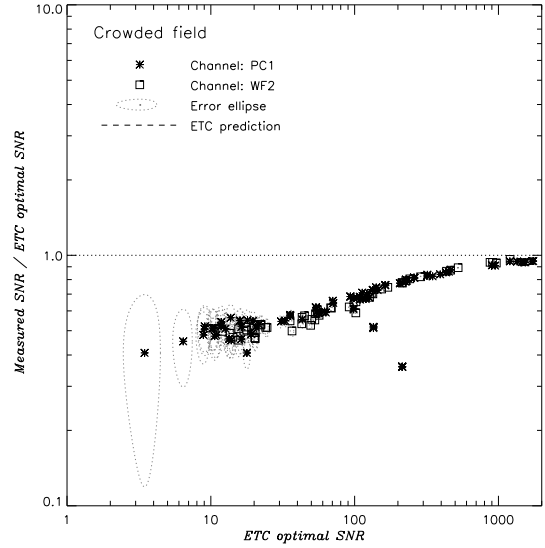}
\caption{ The ratio between the measured SNR (SNR$^{\rm D}$) and the
ETC optimal SNR (SNR$^{\rm P}$) is shown for the PC and the WF2
channels of the WFPC2, in the crowded field case.
\label{fig6}}
\end{figure}

Thus, in this crowded case, it is also apparent that the ETC deviations
are independent of the photometric technique adopted. In sparse fields,
where aperture photometry and PSF fitting are equally effective and
reliable, Figures~\ref{fig3} and~\ref{fig4} already prove that the ETC
predictions depart from the measured data, although by a smaller amount
than that applicable in the crowded case.  Finally, in
Figure~\ref{fig6} we compare the predictions of the ETC with the
actual measurements for both the PC and WF, to show that the behaviour
of the ETC applies regardless of the channel.
 
\begin{figure}
\plotone{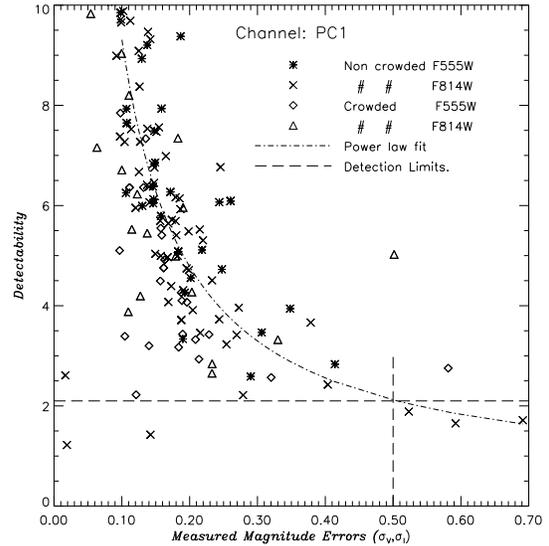}
\caption{ Detectability $d$ versus measured magnitude error
 ($\sigma_m$). An uncertainty $\sigma_m = 0.5 mag$, usually the highest
allowed in most photometric works, corresponds to a detectability $d =
2.1$.
\label{fig7}}
\end{figure}

\begin{figure}
\plotone{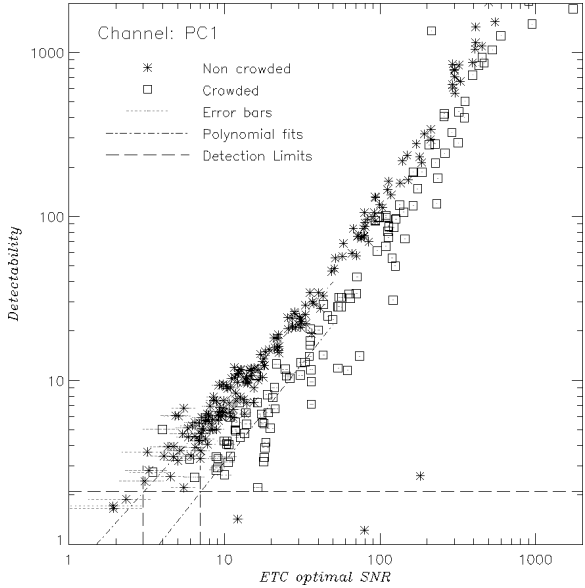}
\caption{ Detectability $d$ versus ETC optimal SNR (SNR$^{\rm P}$). A
value of $d = 2.1$ corresponds to a detection limit of $SNR^P = 3.0$ or
$SNR^P = 7.0$ respectively for the sparse and the crowded case.
 \label{fig8}}
\end{figure}

The relevance of the above considerations becomes clear when one uses
an ETC to simulate very deep observations, especially when a comparison
between instruments, e.g. ACS/WFC and WFPC2, is required to compare the
limiting magnitude in given exposure times. As experience shows, a star
finding programme is able to detect a faint point source only when its
brightest pixel is at least $2$ or $3 \, \sigma_{\rm sky}$ above the
sky background (where $\sigma_{\rm sky}$ is the standard deviation of
the background), with  a value of $\sim 5$ or more being the typical
prerequisite in most faint photometry precision applications. If we
plot the so called object {\it detectability} $d$, defined as:

\begin{equation}
\label{eq3}
d = \frac{\rm Peak - Sky}{\sigma_{\rm sky}}
\end{equation}

as a function of the magnitude error $\sigma_m$, we obtain the graph
in Figure~\ref{fig7}. Here we notice that the detectability (which is
practically independent of the filter and crowding) drops to the value
of $d = 2.1$ just when the magnitude error approaches $0.5\,mag$, which
is usually considered the maximum allowed error in canonical
photometric work. By relating the detectability $d$ with the ETC
optimal SNR, ${\rm SNR^P}$, as done in Figure~\ref{fig8}, we see that
$d = 2.1$ corresponds to an ETC optimal SNR of $3.0$ for the non
crowded case and to $7.0$ for the crowded case. This literally means
that if we need to know the magnitude of the faintest detectable star
in an observation of a stellar field with the WFPC2 we should query the
ETC, setting ``average sky'', for a SNR of $7.0$ and $3.0$, respectively
in a crowded and in a sparse environment. It is normally assumed that a
$3\,\sigma$ detection requires a SNR of 3, but in the case of the SNR
provided by the WFPC2 ETC, this is only true for an isolated
object.

\section{Discussion and corrections}

The direct consequence of what we have illustrated so far is that, if
the ETC were used to plan observations of faint stars in a globular
cluster like M\,4 with the WFPC2, the predicted exposure time could be
considerably underestimated. Conversely, the same predictions
would be almost correct for a star of equal brightness in a sparse field. In
the following we try and provide an empirical correction formula that
can be applied to the SNR given by the WFPC2 ETC to compensate for the
effects of crowding.

\begin{figure}
\plotone{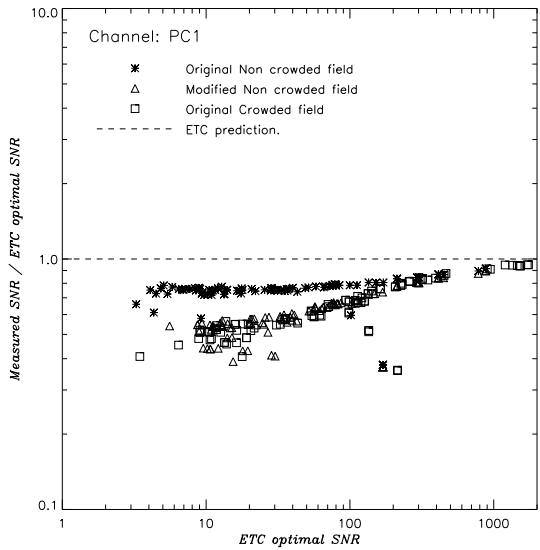}
\caption{ The ratio between the measured SNR (SNR$^{\rm D}$) and the
ETC optimal SNR (SNR$^{\rm P}$) is shown for the crowded and sparse
fields, before and after the artificial brightening of the sparse field
background.
\label{fig9}}
\end{figure}

In order to understand the discrepancy between the expected and
measured SNR and to clarify how to exactly account for the effects of
crowding in the simulations, we artificially modified the background
level and photon noise in the sparse field so as to reproduce the sky
level and sky variance measured on the crowded field.  In practice, we
added to the sparse field a Gaussian noise with a mean equal to the
difference in the sky level between the two fields and a variance equal
to the quadratic difference of the sky variances between them.  The SNR
diagramme for the modified image (Figure~\ref{fig9}) reveals that the
locus of the modified sparse data points shifts towards and perfectly
overlaps the crowded field locus.  This tells us, as expected, that the
increased background level resulting from crowding is responsible for
the differences shown in Figures~\ref{fig3} and~\ref{fig4} between
sparse and crowded fields.

\begin{figure}
\plotone{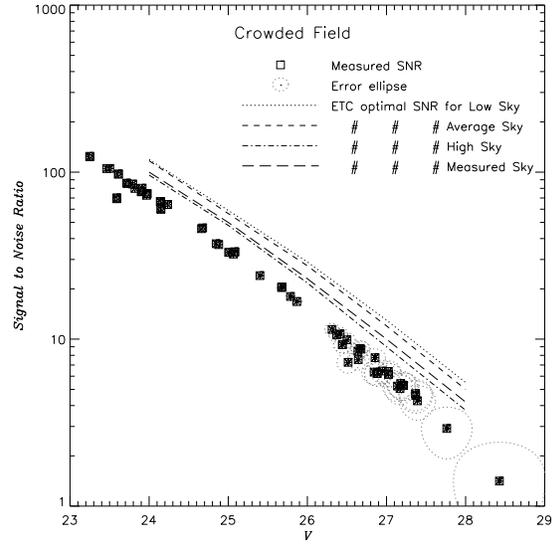}
\caption{ Comparison between the measured SNR (SNR$^{\rm D}$) and the ETC
predictions for i) default low background, ii) default average background, iii) default high background and iv) actually measured background. \label{fig10}}
\end{figure}

\begin{figure}
\plotone{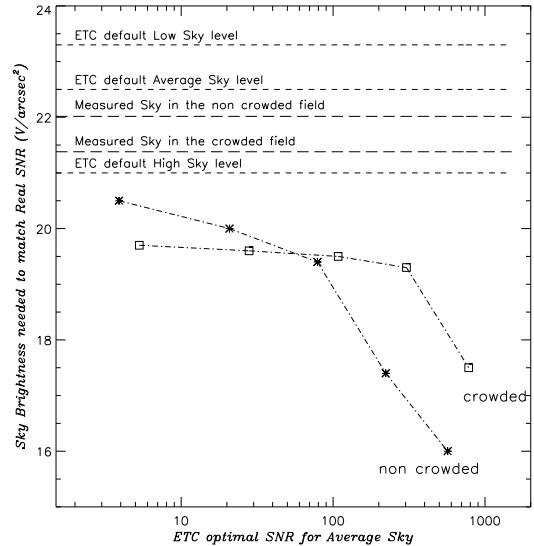}
\caption{ Values to enter in the User Specified Sky Background
parameter of the WFPC2 ETC in order to force the ETC to match the
measured SNR for crowded and non crowded fields.
 \label{fig11}}
\end{figure}

It is, however, true that the ETC gives the SNR under the best possible
sky conditions, which are rarely encountered, if ever, in real
observations.  Moreover, it is generally not expected of the ETC to
take account of the position and brightness of all the stars in the
field as would be necessary to simulate how crowding increases the
background level. We have, therefore, manually set the ETC sky
brightness to match the levels directly measured with the DAOPhot SKY
task on the crowded image (i.e.. the mode of the levels distribution),
hoping in this way to force the SNR simulated by the ETC to agree with
our measurements. In fact, the results change only marginally, as shown
in Figure\,\ref{fig10}, where ${\rm SNR^D}$ and ${\rm SNR^P}$ are
plotted against the observed magnitude (Johnson $V$ in this case).  The
ETC simulation gets closer to the real data, but it does not still
match them. Moreover, it seems as if a suitable value for the
background cannot be found at all as shown in Figure~\ref{fig11}, where
one sees that the sky value that would force the ETC prediction to
match ${\rm SNR^D}$, changes significantly as a function of star
brightness.

We must, thus, conclude that the treatment of the background is a major
issue for the WFPC2 ETC, although that alone cannot explain the whole
discrepancy. It goes without saying that we have verified and confirmed
that the predictions of the ETC as concerns the count rates per pixel
in the source and background are precise to within an accuracy of
10\,\%, as one would expect of a professional tool. We have also
repeated all our tests on the individual frames, compared in turn with
the predictions of the ETC for a case of CR-SPLIT=1. The result being
the same, we can exclude an error in either the way in which we
combined the data or in the way in which the ETC accounts for
CR-SPLIT$> 1$.  The rest of the discrepancy, then, must be attributed
to the way in which the noise is estimated, the signal being correct. A
delicate issue could be, for instance, the value and operational
definition of $\sigma_{\rm sky}$. We notice here that large variations
in the value of $\sigma_{\rm sky}$ are possible, in the crowded
environment, depending as to whether we measure it with the IRAF SKY
task, which fits a Gaussian around the mode, or as the standard
deviation that one obtains by manual analysis over the darkest regions
of the background in the image. In fact the latter can be up to 3 times
smaller than the former, and also 2 times smaller than the mean sigma
as measured inside the photometric sky annulus around each star.
Conversely, all these numbers turn out to be quite similar for the
sparse field image.

To try and account for the possible sources of the residual error, we
considered recent results published by \citet{ber01}, who uses Fourier
analysis and Fisher information matrices to show to which extent the
SNR of a point source depends on factors which normally are not
considered in ETC programmes, such as pixel size, intra-pixel response
function, extra-pixel charge diffusion and cosmic ray hits.

According to this work, a programme that does not take all these
parameters into account may overestimate the SNR by up to a factor of\,2.
More precisely, whenever background limited point source photometry
is involved, the key factor for the SNR calculation, namely the
``effective area'' $A_{SN}$ (see equation\,12 in Bernstein 2001),
strongly depends on the detector geometry, such as pixel size,
under-sampling factor, intra-pixel response function and charge
diffusion. The finite pixel size plays an important role, as even a
Nyquist sampled pixel (i.e. one $\lambda /2D$ in size) causes a 13\,\%
degradation in the SNR of a faint star and the same applies to
extra-pixel charge diffusion.

\begin{figure}
\plotone{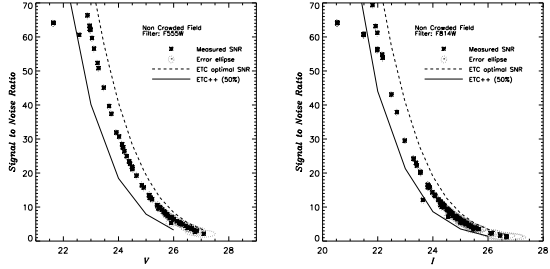}
\caption{ Comparison between the prediction of the WFPC2 ETC v.3.0,
the prediction of the ETC++ software and the measured SNR (SNR$^{\rm D}$),
for the crowded and non crowded cases in the two filters (both ETCs
were used here after setting the sky magnitude to the value measured
in the real images). \label{fig12}}
\end{figure}

In order to check whether these problems also affect the WFPC2, we
configured Bernstein's ``ETC++'' software to simulate WFPC2 point source
photometry for the sparse field. The result is shown in
Figure~\ref{fig12} where the measured SNR (${\rm SNR^D}$), the ETC
optimal SNR (${\rm SNR^P}$) and the ETC++ SNR for aperture photometry
are plotted against the stellar magnitude.  The ETC++ gives a
confidence level for its results as the value of the cumulative
function of the stars distribution above the computed SNR. The ETC++
line in Figure~\ref{fig12} means that 50\,\% of the stars of any given
magnitude should be above this line.  The WFPC2 ETC does not give
confidence levels, but we can assume that its SNR is computed as the
mean of the SNR distribution at any given magnitude, i.e. at 50\,\%
confidence level.  If this is the case, Figure~\ref{fig12} indicates
that the actual SNR is located in between the WFPC2 ETC and the ETC++
predictions, thus confirming the difficulty of any analytical ETC in
reliably estimating the SNR.

Thus, a correction for the currently on-line WFPC2 ETC can only be
empirical in nature. The following formula can be used to obtain a
realistic estimate of the SNR:

\begin{equation}
\label{eq4}
{\rm SNR^C} \simeq (60 \cdot C + 17) \cdot (e^{-0.012 \cdot {\rm SNR^P}}-1)
+ 0.93 \cdot {\rm SNR^P}
\end{equation}

\noindent
where ${\rm SNR^P}$ is the SNR estimated by the ETC without correction
for aperture photometry and $C$ is a measure of the crowding, defined
as the logarithm of the ratio between the total area of the chip and
the number of pixel with value lower than the modal sky value plus one
standard deviation.  For example $C$ is equal to $0.05$ for our sparse
field, whereas it grows to $0.42$ in the crowded case of M\,4. For
faint stars, e.g. for ${\rm SNR^P} \lesssim 20$, this equation can be
roughly approximated by the rule of thumb that the actual SNR is about
$1/2$, or $2/3$, of the ${\rm SNR^P}$, respectively for a crowded and
non crowded environment.  It should be noted that not even in an ideal
case of zero crowding ($C \simeq 0$) would the measured SNR match the
prediction of the ETC, since there would still be a discrepancy of the
same order of that found in the sparse case.

The advantage of this formula is that ${\rm SNR^C} = 3$ would now
always imply a $3\,\sigma$ detection, regardless of the level of
crowding in the image. The correction that we propose would allow an
observer to accurately plan his observations and make the best use of
the HST time. For the low SNR regime (e.g. ${\rm SNR^P} \lesssim 50$), equation\,\ref{eq4} can actually be rewritten to more
explicitly show the effects of crowding on the exposure time:

\begin{equation}
\label{eq5}
{\rm t^C} \simeq {\rm t^P} \cdot \frac{{\rm SNR^P}}
{(60 \cdot C + 17) \cdot (e^{-0.012 \cdot {\rm SNR^P}}-1)
+ 0.93 \cdot {\rm SNR^P}}
\end{equation}

\noindent
where ${\rm t^P}$ is the exposure time predicted by the ETC to reach a
certain SNR and ${\rm t^C}$ is its actual value.

\begin{figure}
\plotone{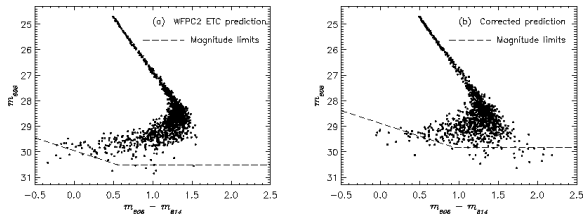}
\caption{ Comparison between (a) the WFPC2 ETC predictions (SNR$^{\rm
P}$) and (b) our correction of equation\,\ref{eq4} (SNR$^{\rm C}$), in
a simulation of a 120 HST orbits observation of the white dwarfs
cooling sequence in NGC6397, in a colour--magnitude diagramme made
through the filters F606W and F814W.
\label{fig13}}
\end{figure}

An example of how serious the underestimate of the exposure time can be
when the ETC is not used with the above caveat in mind is given in
Figures~\ref{fig13}a and \ref{fig13}b for a crowded environment. There
we show a simulation of the detectability of the white dwarf cooling
sequence with the WFPC2 in NGC6397, the nearest globular cluster,
through the filters F606W and F814W. We have adopted the theoretical WD
cooling sequence of \citet{pm02} which provide a perfectly thin
isochrone and have applied to it the colour and magnitude uncertainty
that one obtains from the estimated SNR by inverting
equation\,\ref{eq1}.  Two cases are shown:  one (a) as predicted by the
WFPC2 ETC and one (b) for our corrected estimate of
equation\,\ref{eq4}.  The difference is outstanding, as the ETC
predictions, taken at face value and ignoring the effects of crowding,
would suggest that the sequence is not spread very much by photometric
errors and its quasi-horizontal tail between $m_{606} = 29$ and
$m_{606} = 30$ is clearly noticeable, whereas in our realistic
simulation the sequence is widely spread and its lower part lies well
below the detection limit.
 
The delicacy of the issue is immediately apparent when one considers
that, based on the ETC estimates, one would deem that 120 orbits are
sufficient to reliably secure the white dwarf cooling sequence in the
colour--magnitude diagramme of NGC6397 down to $m_{606} = 30.5$ and
$m_{814} = 30$, whereas, in fact, the correction shows that as many as
255 orbits would be needed to comfortably reach those limits with the
WFPC2.

All of the above considerations are valid not only for the WFPC2, but
also for any analytical ETC in general, especially when used to
estimate the SNR of stars embedded in crowded environment or when the
detector considerably under-samples the PSF, as suggested in
\citet{ber01}. We should underline here, however, that this does not
mean at all that the ETCs are unreliable nor that they are useless. One
of the most important and practical reasons for having a standardised
ETC is to allow the telescope time allocation committees to compare all
the proposals on equal footing. In this respect, the ETC does not
necessarily need to be accurate. Clearly, the better the detector's
cosmetics, intra-pixel response, charge diffusion and readout noise,
the closer will the real photometry be to the ETC prediction. Thus, we
expect, for example, a better behaviour of the ACS/WFC on-line
simulator with respect to the WFPC2. 

A non-analytical SNR calculator, which would simulate the whole
observing session, including the dithering pattern, by numerically
reproducing the real field (i.e. with the correct stellar positions and
brightness, as imaged by a realistic model of the detector) and which
uses the same photometric tools that will be adopted by the user (such
as DAOPhot, ALLSTAR, and the like), would be, in our opinion, the best
method to accurately predict the expected performances of any planned
observing programme providing reliable results. Alternatively, at least
for imaging ETCs which have very few configuration parameters and are
fairly stable such as those in space telescopes, one should consider
empirical modeling. One can take real results, such as we did in this
paper, to calibrate an ETC which in turn interpolates between
calibrations. In this way, the use of an empirical correction formula
such as the one proposed here would guarantee a closer matching between
simulations and real observations.

\section{Summary and conclusions}

The results of the WFPC2 exposure time calculator for point sources
have been analysed by direct comparison with aperture and PSF
photometry on real archival images. Significant deviations have been
found between the ETC predictions and the actual photometry on the real
data.  Specifically, the analysis shows that the ETC deviations are
{\it i)} independent of the filter, {\it ii)} independent of the choice
of optimised aperture photometry or PFS fitting photometry, {\it iii)}
independent of the PC or WF channel used, {\it iv)} strongly dependent
upon the level of crowding in the field and that {\it v)} the ETC
systematically overestimates the SNR, slightly for the bright sources
and more seriously for faint sources close to the detection limit.
Moreover, when data reduction follows the optimised aperture photometry
method, the measured SNR will be as good as that obtained with PSF
fitting and there is no need to apply the aperture photometry
conversion suggested in the ETC documentation. An empirical correction
formula is given to compute realistic SNR estimates, so as to assist
observation planning when extremely faint sources have to be imaged, an
example of which is presented. Manually increasing the value of the sky
brightness in the simulator, so as to mimic the effects of crowding,
shows that, although important, the background level is not the key
parameter to explain the discrepancy, which is present even for data
collected in rather sparse environments. Thus, it is not possible to
correct the WFPC2 ETC predictions by just modifying the sky level.  A
comparison with a software tool developped by \citet{ber01}, whose
predictions slightly underestimate the SNR at variance with the WFPC2
ETC, suggests that the effects of pixel size, charge diffusion and
cosmic rays hits could be more important than previously thought.

\acknowledgments

It's our pleasure to thank H. Ferguson, M. Stiavelli, S. Casertano, F.
Massi, L. Pulone and R. Buonanno for helpful discussions. We are
indebted to F. Valdes, the referee of this paper, for his useful
comments and suggestions. G. Li Causi is particularly grateful to the
ESO Director General's Discretionary Fund for supporting his work. We
also wish to thank Gary Bernstein for making his ETC++ software
available to us.

\clearpage

FIGURE CAPTIONS:
\\
\\

Figure 1: Negative image of a crowded field in the globular cluster M4
obtained with the PC channel of the WFPC2 through the F555W filter
\citep{rich97}.

Figure 2: Negative image of a sparse field obtained with the PC
\citep{iba98}.

Figure 3: The ratio between the SNR measured in crowded and sparse
fields (SNR$^{\rm D}$ in the text) and the WFPC2 ETC prediction for
aperture photometry (SNR$^{\rm A}$ in the text) is shown for the F555W
and F814W filters.

Figure 4: The ratio between the SNR measured in crowded and sparse
fields (SNR$^{\rm D}$ in the text) and the WFPC2 ETC prediction for
PSF-fitting photometry (SNR$^{\rm P}$ in the text) is shown for the
F555W and F814W filters.  The right hand side axis applies to the low
SNR regime ($\lesssim 50$) and indicates the amount of the time
estimation error, i.e. the ratio between the actual exposure time (Equation\,7 in the text) and that estimated by the ETC, for a given SNR in the abscissa.

Figure 5: Measured magnitude error from PSF-fitting photometry of
\citet{iba99} ($\sigma_m^P$) and from our optimized aperture photometry
($\sigma_m^A$), as a function of the magnitude, compared with the WFPC2
ETC prediction, for the crowded field and F555W filter.

Figure 6: The ratio between the measured SNR (SNR$^{\rm D}$) and the
ETC optimal SNR (SNR$^{\rm P}$) is shown for the PC and the WF2
channels of the WFPC2, in the crowded field case.

Figure 7: Detectability $d$ versus measured magnitude error
($\sigma_m$). An uncertainty $\sigma_m = 0.5 mag$, usually the highest
allowed in most photometric works, corresponds to a detectability $d =
2.1$.

Figure 8: Detectability $d$ versus ETC optimal SNR (SNR$^{\rm P}$). A
value of $d = 2.1$ corresponds to a detection limit of $SNR^P = 3.0$ or
$SNR^P = 7.0$ respectively for the sparse and the crowded case.

Figure 9: The ratio between the measured SNR (SNR$^{\rm D}$) and the
ETC optimal SNR (SNR$^{\rm P}$) is shown for the crowded and sparse
fields, before and after the artificial brightening of the sparse field
background.

Figure 10: Comparison between the measured SNR (SNR$^{\rm D}$) and the ETC
predictions for i) default low background, ii) default average background, iii) default high background and iv) actually measured background.

Figure 11: Values to enter in the User Specified Sky Background
parameter of the WFPC2 ETC in order to force the ETC to match the
measured SNR for crowded and non crowded fields.

Figure 12: Comparison between the prediction of the WFPC2 ETC v.3.0,
the prediction of the ETC++ software and the measured SNR (SNR$^{\rm
D}$), for the crowded and non crowded cases in the two filters (both
ETCs were used here after setting the sky magnitude to the value
measured in the real images).

Figure 13: Comparison between (a) the WFPC2 ETC predictions (SNR$^{\rm
P}$) and (b) our correction of equation\,\ref{eq4} (SNR$^{\rm C}$), in
a simulation of a 120 HST orbits observation of the white dwarfs
cooling sequence in NGC6397, in a colour--magnitude diagramme made
through the filters F606W and F814W.

\clearpage

\end{document}